\RequirePackage{fix-cm}
\documentclass[smallextended]{svjour3}       
\smartqed  
\usepackage{graphicx}
\usepackage{multirow}
\begin{document}

\title{Selective Information Passing for MR/CT Image Segmentation
}


\author{Qikui Zhu \and
        Liang Li\textsuperscript{\#} \and
        Jiangnan Hao\textsuperscript{\#}\and
        Yunfei Zha\textsuperscript{\#}\and
        Yan Zhang\textsuperscript{\#} \and
        Yanxiang Cheng\textsuperscript{\#} \and
        Fei Liao\textsuperscript{\#} \and
        Pingxiang Li\textsuperscript{\#}
}


\institute{Qikui Zhu \at
              School of Computer Science, Wuhan University, Wuhan, China. \email{QikuiZhu@whu.edu.cn}           
           \and
           Liang Li \at
           Department of Radiology, Renmin Hospital of Wuhan University, Wuhan, China. \email{liliang\_082@163.com}
           \and
           Jiangnan Hao \at
           Xi'an Aeronautical University. \email{ haojiangnan@mail.nwpu.edu.cn}
           \and
           Yunfei Zha \at
           Renmin Hospital of Wuhan University, Wuhan, Hubei Province, China. \email{zhayunfei999@126.com}
           \and
           Yan Zhang \at
           Department of Clinical Laboratory, Renmin Hospital of Wuhan University, Wuhan, China. \email{peneyyan@mail.ustc.edu.cn}
           \and
           Yanxiang Cheng \at
           Department of Obstetrics and Gynecology, Renmin Hospital of Wuhan University, Wuhan, China. \email{yanxiangcheng@whu.edu.cn}
           \and
           Fei Liao \at
           Department of Gastroenterology, Renmin Hospital of Wuhan University, Wuhan, China. \email{feiliao@whu.edu.cn}
           \and
           Pingxiang Li \at
           Renmin Hospital of Wuhan University, Wuhan, China. \email{pxli@whu.edu.cn}
           \and
'\#' indicates co-corresponding authors.
}

\date{Received: date / Accepted: date}

\maketitle

\begin{abstract}
Automated medical image segmentation plays an important role in many clinical applications, which however is a very challenging task, due to complex background texture, lack of clear boundary and significant shape and texture variation between images. Many researchers proposed an encoder-decoder architecture with skip connections to combine low-level feature maps from the encoder path with high-level feature maps from the decoder path for automatically segmenting medical images. The skip connections have been shown to be effective in recovering fine-grained details of the target objects and may facilitate the gradient back-propagation. However, not all the feature maps transmitted by those connections contribute positively to the network performance. In this paper, to adaptively select useful information to pass through those skip connections, we propose a novel 3D network with self-supervised function, named selective information passing network (SIP-Net). We evaluate our proposed model on the MICCAI Prostate MR Image Segmentation 2012 Grant Challenge dataset, TCIA Pancreas CT-82 and MICCAI 2017 Liver Tumor Segmentation (LiTS) Challenge dataset. The experimental results across these data sets show that our model achieved improved segmentation results and outperformed other state-of-the-art methods. The source code of this work is available at https://github.com/ahukui/SIPNet.
\end{abstract}

\section{Introduction}
\begin{figure}
  \centering
  \includegraphics[width=0.7\columnwidth]{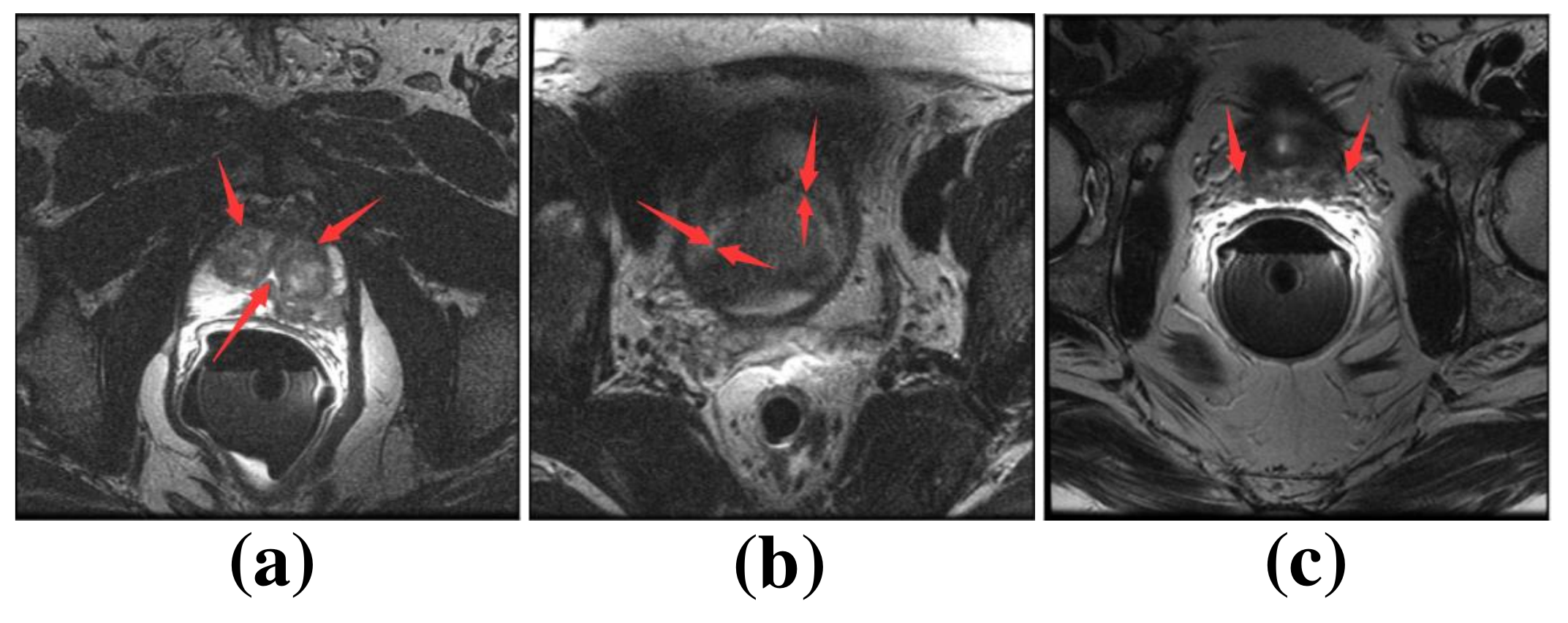}
  \caption{Challenges in segmenting the prostate from MR images. (a) Noise inside prostate. (b) Weak boundary. (c) Surrounding tissues having similar intensity distribution with prostate.}\label{fig:challenges}
\end{figure}

Medical image segmentation is an essential part of medical image analysis.
Accurate segmentation of medical image provides very useful information for computer aided diagnosis and treatment of cancers as well as other diseases\cite{zhu2019multi}. For instance, segmentation of the liver and tumors plays an important role in hepatocellular carcinoma diagnosis \cite{heimann2009comparison}. Accurate prostate segmentation is useful for treatment planning and therapeutic procedures for prostate cancer\cite{liao2013representation,zhu2019boundary,zhu2018exploiting}.
However, automated medical image segmentation is very challenging for several reasons. Taking prostate segmentation as an example: First, due to many slices only have small part of segmented tissues specifically at the apex and base, which always led to those slices lack of clear boundary and make the automated segmentation fail. Second, imaging artifacts always distribute in the whole image randomly, which negatively influence the process of segmentation. Third, tissues can have a wide variation in size and shape among different slices, which adds to the complexity of segmentation. Fourth, the complex background and fuzzy boundary also make the segmentation process challenging. Furthermore, different from natural images dataset, the size of available medical image dataset is limited.
Fig.\ref{fig:challenges} shows examples of prostate MR images.
Fig.\ref{fig:challenges}(a) shows the phenomenon that imaging artifacts locate in prostate region. Fig.\ref{fig:challenges}(b) shows prostate region lacks clear boundary. Fig.\ref{fig:challenges}(c) shows the prostate and surrounding tissues have similar intensity distribution. All of above phenomena bring challenges for automated medical image segmentation.

To overcome the above challenges, over the past few decades, various methods have been developed for medical image segmentation, including machine learning based methods \cite{du2019improved,wu2014multi,bi2019early,li2020robust,wu2014bag,wang2019domain,ijcai2020,wu2013artificial}, level sets\cite{qin2014adaptive}, atlas-based methods\cite{8085166,mcintosh2016contextual,zhu2016metric}, super-pixel segmentation\cite{gao2014multi} and active shape model\cite{yan2011adaptively,7093141}. Recently, deep convolutional neural networks (CNNs) have become the dominant machine learning approach due to their superior performance.
CNNs have achieved state-of-the-art performances in many fields including computer vision \cite{luo2020dimensionality,wu2013artificial,lu2018hierarchical,wu2014boosting,li2019iterative,luo2019sparse}, natural language processing (NLP)\cite{lai2015recurrent,goldberg2016primer,johnson2015semi}, medical image analysis\cite{kim2019multi}. The superiority of CNNs\cite{wu2013multi} can be partially attributed to the ability of learning hierarchical representation of the data.

However, medical image segmentation has a higher-level requirement of accuracy than natural image segmentation, where many excellent networks, such as VGG \cite{he2016deep} and FCN \cite{long2015fully}, cannot be directly utilized. To obtain accurate segmentation results and overcome the problems specific to medical imaging, specific models have been proposed for medical image analysis. For instance, Milletari et al.~\cite{milletari2016v} proposed a network architecture based on the volumetric CNNs, which can segment prostate volumes in a fast and accurate manner. Yu et al.~\cite{yu2017volumetric} proposed a novel volumetric CNN with mixed long and short residual connections for automated prostate segmentation. Gibson et al. \cite{gibson2018automatic} proposed a network called DenseVNet, which can segment the pancreas, esophagus, stomach, liver, spleen, gallbladder, left kidney and duodenum accurately. Li et al.~\cite{li2018h} proposed a novel hybrid densely connected U-Net for liver and tumor segmentation. One thing that these medical image segmentation networks have in common is an encode and decode architecture with skip connections for combining low-level feature maps from the encoder path with high-level feature maps from the decoder path. There is no doubt that the skip connections are effective in recovering fine-grained details of the target objects and help the gradient back-propagation. However, as a lot of information can be passed through those skip connections, do all the feature maps transmitted by those connections always contribute positively to the network performance?

\begin{figure}
  \centering
  \includegraphics[width=0.7\columnwidth]{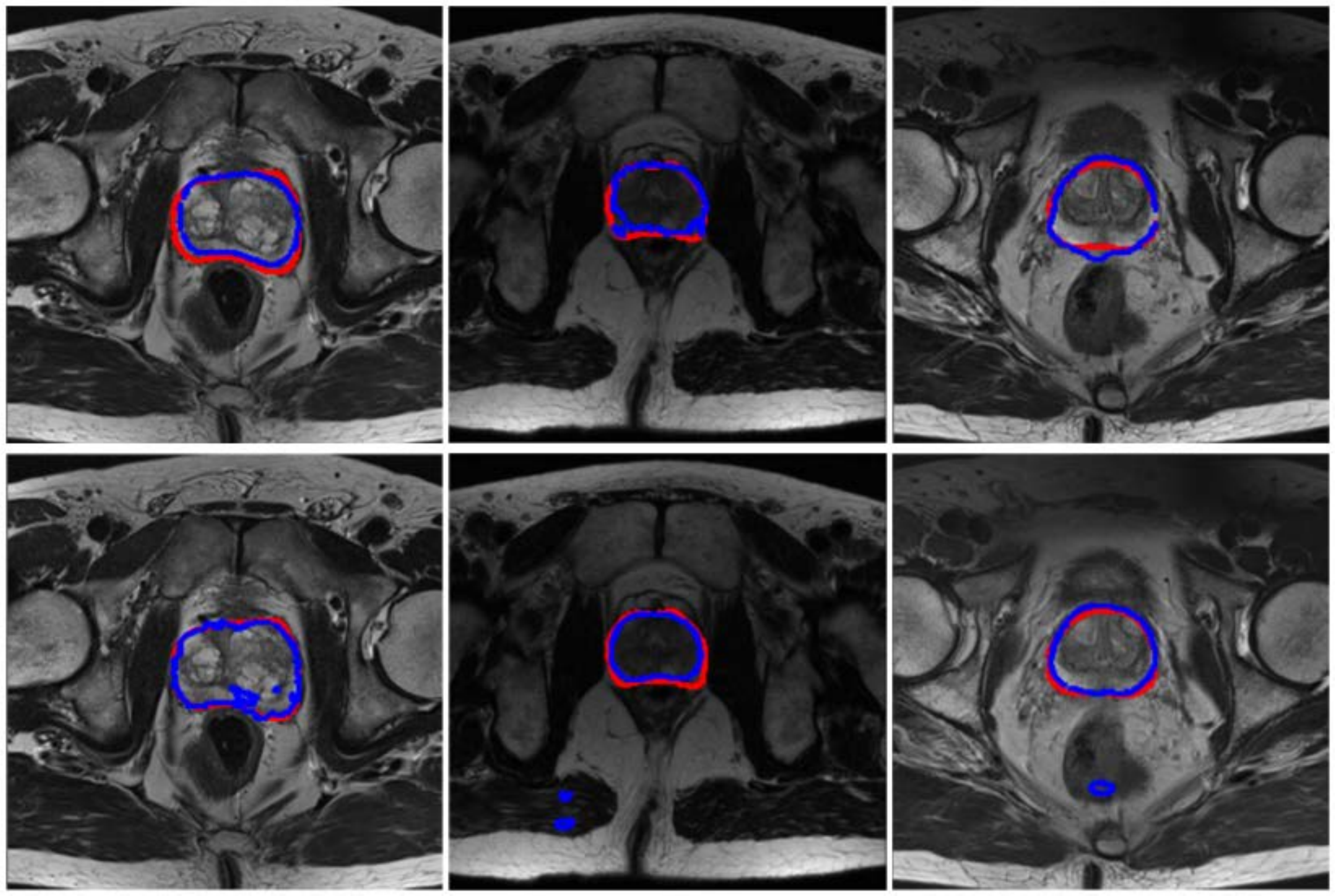}
\caption{Top row: Segmentation results of U-Net without long skip connections, which is in fact equivalent to the original FCN; Bottom row: Segmentation results of U-Net. The red and blue contours indicate the ground truth and segmentation results, respectively.} \label{fig:UnetCompare}
\end{figure}

To answer this question, we analyzed the behavior of the classical U-Net \cite{ronneberger2015u} with and without the long skip connections on the task of prostate segmentation. The segmentation results are shown in Fig.~\ref{fig:UnetCompare}.
Compared with ground truth segmentation, U-Net can obtain finer details and higher accuracy in general. However, the segmentation result of fully convolutional network (FCN) \cite{long2015fully} is smoother and that of U-Net picks up non-prostate regions when those areas are highly inhomogeneous.
%
To make the long skip connections inside the network select the useful information and further improve medical image segmentation performance, in this paper, we propose a novel 3D convolutional network, named SIP-Net. Our proposed SIP-Net adopts Densely-connected Residual Blocks (DRBs) and Attention-focused Modules (AMs). The contributions of this work are summarized as follows.
\begin{itemize}
	
\item
Inspired by the attention mechanism, we propose to integrate attention-focused modules into our model to make the long connections transmit mainly useful features and reduce the negative impact of noise from feature maps. That makes the long connections focus more on the regions to be segmented and the irrelevant noise features from the background and surrounding tissues may be suppressed during feature transmission.

\item
In the same time, to overcome the problem of small size of medical image data, we integrate three different types of connections seamlessly into our proposed model. Together with the above attention-focused modules, these connections improve training efficiency and feature extraction capability of the network by enhancing information propagation and encouraging feature reuse.

\item
To reduce the computational load and more importantly the number of network parameters for alleviating the potential overfitting problem, we design a modified dense block to construct deeper network, which possesses more than 90 convolutional layers but fewer parameters. Our experimental results show that the proposed model is effective in addressing the problems of complex background, fuzzy boundary and large shape variations.
\end{itemize}

The remainder of the paper is organized as follows. Section~\ref{sec:related_works} provides a brief survey of related works. Section~\ref{sec:methods} describes the details of the proposed 3D segmentation network model. In Section~\ref{sec:experiments}, various experiments of segmenting prostate MR images, pancreas CT images, liver CT images are performed to validate the proposed model. The performance of the proposed method is further discussed through ablation studies in Section~\ref{sec:discussions}. Finally, several concluding remarks are drawn in Section~\ref{sec:conclusions}.

\section{Related Works}
\label{sec:related_works}

In this section, we give a brief review of deep learning techniques for semantic image segmentation. We first review the methods for natural image segmentation and then discuss the ones specialized for medical image segmentation.

\subsection{Deep Learning for Semantic Segmentation}

Semantic segmentation is a critical component in image understanding. The task of semantic segmentation is to assign a categorical label to every pixel in an image. Over the past few years, deep learning based methods and in particular convolutional networks (CNNs) have improved segmentation results remarkably in pixel-wise semantic segmentation tasks. This success can be attributed to the ability of hierarchical representation of CNNs.
Fully convolutional networks (FCNs) mark a major milestone in CNN based semantic segmentation \cite{long2015fully}, which is trained end-to-end to perform pixels-to-pixels segmentation.
Since then, FCNs have dominated the field of semantic image segmentation with a number of extensions. For instance, Li et al.~\cite{li2016fully} extended the FCN model for instance-aware semantic segmentation. The model significantly improves the segmentation performance in both accuracy and efficiency.

In the same time, researchers develop deeper and more powerful CNN models to extract more discriminating and complex representation features. For example, Simonyan et al.~\cite{simonyan2014very} proposed a 19-layer network, the famous VGG-19 model, to investigate the effect of the depth of CNNs on their accuracy in large-scale image recognition. He et al.~\cite{he2016deep} presented a residual learning framework to ease the training of very deep networks. Based on this framework, the author proposed a 101-layer model (ResNet-101) and a 152-layer model (ResNet-152) and won the first places in several tracks in ILSVRC \& COCO 2015 competitions\footnote{http://image-net.org/challenges/ilsvrc+mscoco2015}. Soon after that, Wu et al.~\cite{wu2016high} proposed a method for high-performance semantic image segmentation based on the deep residual networks, which achieves the state-of-the-art performance.

\subsection{Deep Learning for Medical Image Segmentation}

Recently, deep CNNs have also become the dominant approach for medical image segmentation. Many researchers have employed various CNN models to segment images from different medical imaging modalities. In our previous work, we proposed a deeply supervised CNN model \cite{zhu2017deeply}, which employs additional supervised layers and utilizes the residual information to segment the prostate from MR image.
To exploit the information from different views of volumetric images but without using 3D convolutions, Mortazi et al.~\cite{mortazi2017cardiacnet} proposed a multi-view CNN to segment structures from cardiac MR images by using an adaptive fusion strategy. Han~\cite{han2017automatic} proposed a 2.5D model to segment liver tumors, which takes a stack of adjacent slices as input and produces the segmentation map corresponding to the center slice. 

To fully exploit 3D spatial information in volumetric MR images, a few studies employed 3D convolutions to build CNNs. For example, Li et al.~\cite{li2018h} proposed a novel hybrid densely connected U-Net for liver and tumor segmentation. The proposed model consists of a 2D Dense-U-Net and a 3D counterpart, which can extract intra-slice features and hierarchically aggregate 3D contexts under the spirit of the auto-context algorithm \cite{tu2008auto}.
Chen et al.~\cite{chen2016voxresnet} extended deep residual learning into a 3D for 3D brain segmentation. This model also seamlessly integrates the low-level image appearance features, implicit shape information and high-level context together for further improving the 3D segmentation performance.
Recently, Yu et al.~\cite{yu2017automatic} proposed a novel densely-connected volumetric CNN, which adopts the 3D fully convolutional architecture to automatically segment cardiac and vascular structures from 3D cardiac MR images. 

Compared with 2D networks, these 3D networks were able to achieve better segmentation performance.
However, 3D CNNs have a much larger number of parameters and computational complexity than 2D networks. Due to the limited size of typical medical image dataset, it makes the network difficult to train. Furthermore, the trained network easily suffers from overfitting. Therefore, there is still much need in pushing the potential of CNNs by effectively extracting the information from limited training data to improve the segmentation performance and also reduce the complexity of the networks to avoid overfitting.



\begin{figure*}[t]
  \centering
  \includegraphics[width=.8\textwidth]{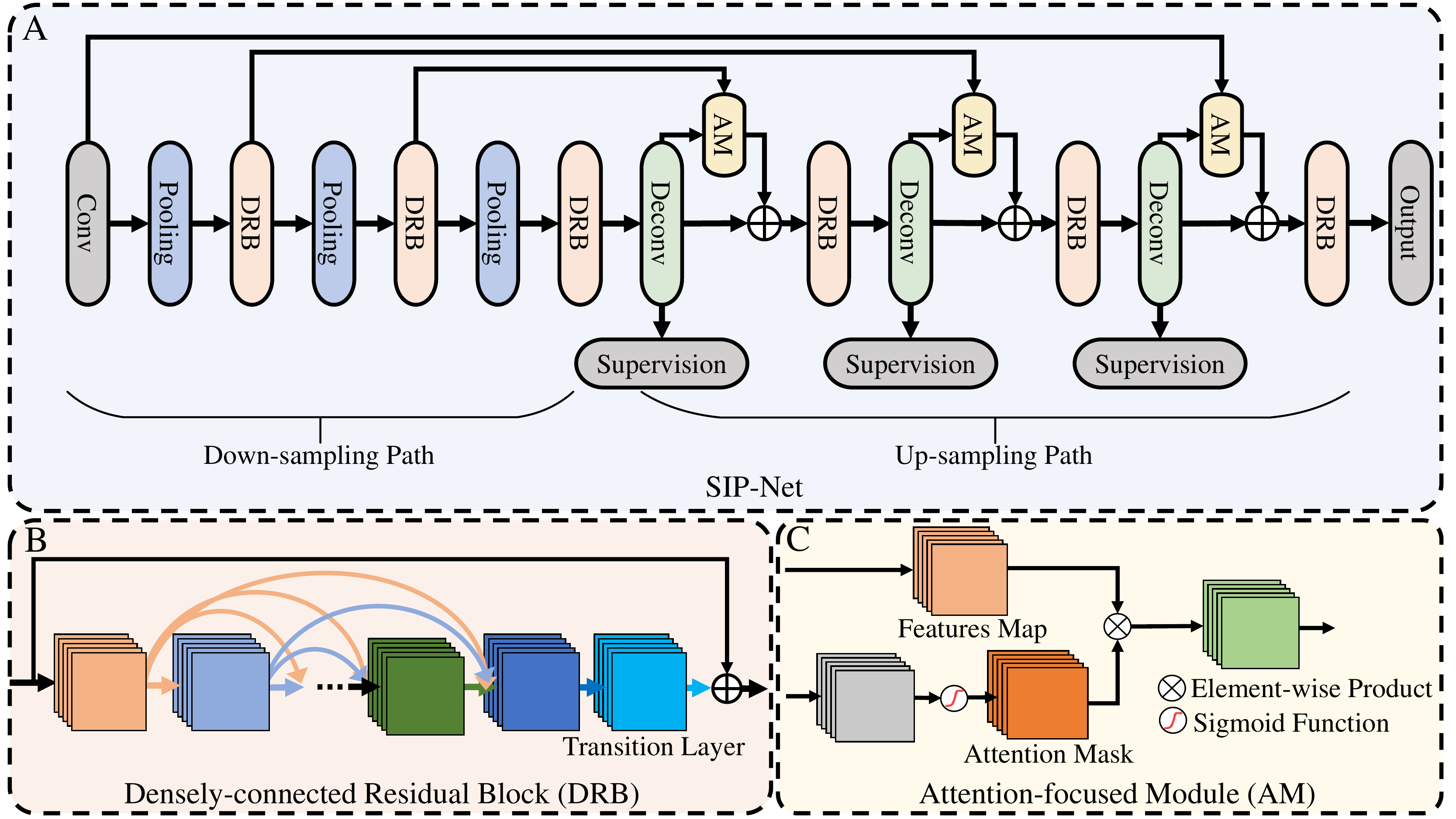}
  \caption{The illustration of the pipeline for medical image segmentation. A. The proposed SIP-Net. B. The structure of Densely-connected Residual Block (DRB). C. The structure of attention-focused module (AM).}\label{fig:network_structure}
\end{figure*}
\section{Methods}
\label{sec:methods}
In this section, we first give an overview of the proposed SIP-Net and then discuss each module of the model in detail.
\subsection{SIP-Net}
In order to fully use the 3D spatial contextual information of volumetric data to accurately segment medical images, in this paper, we propose a 3D CNN with densely-connected residual blocks (DRBs) and attention-focused modules (AMs), named SIP-Net. The overall structure is shown in Fig.~\ref{fig:network_structure}. The proposed SIP-Net contains two paths: down-sampling path and up-sampling path. The down-sampling path consists of one convolutional block, three DRBs and three average pooling layers. The pooling layers use stride of 2, which gradually reduces the resolution of feature map and increases the receptive field of the convolutional layers. 
To obtain accurate segmentation result in the original image resolution, an up-sampling path is implemented, which contains three deconvolutional layers and three DRBs. The deconvolutional layers gradually up-sample the feature maps until reaching the original input size. The overall illustration and detailed structure of proposed network are shown in Fig.~\ref{fig:network_structure}(A) and Table~\ref{tab:structure}, respectively.


In our proposed SIP-Net, we could have used the long connections between the down-sampling path and up-sampling path to connect the blocks in the same resolution level in the down-sampling and up-sampling paths.
However, our study shows that simply adding the long connections may cause noisy segmentation by considering part of noise as shown in Fig.~\ref{fig:UnetCompare}. To make the network focus more on the segmented region and reduce the negative influence from background and surrounding tissues, in this paper, we employ the attention mechanism in our proposed model. Inspired by the attention mechanism in residual attention network~\cite{wang2017residual}, three attention-focused modules are used in up-sampling path, which reduces irrelevant noise in background and surrounding tissues and holds segmenting features from down-sampling path and make the network focus more on the areas to be segmented in the up-sampling path.


In addition, to enforce the attention-focused modules to act effectively as information pass filters, we also integrate a deep supervision mechanism \cite{lee2015deeply} for the attention-focused modules. An additional supervision layer is added after each deconvolutional layer. Each of the three additional supervision layers consists of one up-sampling layer for enlarging the feature map to its original size and one convolutional layer for obtaining the segmentation output as shown in Fig.~\ref{fig:network_structure}(A). Those additional supervision layers bring two advantages. First, it helps to supervise the attention-focused modules to produce accurate attention masks to guide information passing. Second, it can accelerate the network convergence speed during training due to the shorter backpropagation paths from the additional supervision outputs.

In total, our proposed SIP-Net has more than 100 layers in depth including convolutional layers, pooling layers, layers in dense blocks, transitional layers, dropout layers and deconvolutional layers. The dense layers contain different numbers of BN-ReLU-Conv(1$\times$1$\times$1)-BN-ReLU-Conv(3$\times$3$\times$3) with growth rate of $k = 32$. The transition layer is implemented using a BN-ReLU-Conv(1$\times$1$\times$1) layer. After each Conv(3$\times$3$\times$3) layer, a dropout layer with 0.3 dropout rate is added to help deal with the potential overfitting problem. The designs of DRB and AM are shown in Fig.~\ref{fig:network_structure}(B) and Fig.~\ref{fig:network_structure}(C) and the details are given in the rest of this section.




\begin{table}
  \newcommand{\tabincell}[2]{\begin{tabular}{@{}#1@{}}#2\end{tabular}}
  \caption{Detailed structure of the SIP-Net.}
  \centering
  \fontsize{9.6}{14}\selectfont
    \begin{tabular}{c|c|c}
        \hline
        \hline
         &Feature Size & SIP-Net (k=32)\\\cline{1-3}
        \hline         \hline
        Input & 96$\times$96$\times$16$\times$1& \\\cline{1-3}

        Convolution1 & 96$\times$96$\times$16$\times$64&3$\times$3$\times$3 conv\\\cline{1-3}

        Pooling &48$\times$48$\times$8$\times$64& \tabincell{c}{2$\times$2$\times$2 avg. pool\\stride=2}\\\cline{1-3}

        DRB1 &48$\times$48$\times$8$\times$192& \tabincell{c}{1$\times$1$\times$1 conv \\3$\times$3$\times$3 conv\\num=4}\\\cline{1-3}

        TransLayer1 & 48$\times$48$\times$8$\times$128 & 1$\times$1$\times$1 conv\\\cline{1-3}

        Pooling &24$\times$24$\times$4$\times$128&\tabincell{c}{2$\times$2$\times$2 avg. pool \\stride=2}\\\cline{1-3}

        DRB2 &24$\times$24$\times$4$\times$384& \tabincell{c}{ 1$\times$1$\times$1 conv \\3$\times$3$\times$3 conv \\num=8}\\\cline{1-3}

        TransLayer2 & 24$\times$24$\times$4$\times$256 & 1$\times$1$\times$1 conv\\\cline{1-3}

        Pooling &12$\times$12$\times$2$\times$256&\tabincell{c}{2$\times$2$\times$2 avg. pool \\stride=2}\\\cline{1-3}

        DRB3 &12$\times$12$\times$2$\times$768&\tabincell{c}{ 1$\times$1$\times$1 conv \\3$\times$3$\times$3 conv \\num=16}\\\cline{1-3}

        TransLayer3 & 12$\times$12$\times$2$\times$512 & 1$\times$1$\times$1 conv\\\cline{1-3}

        Deconvolution1 &24$\times$24$\times$4$\times$256& \tabincell{c}{3$\times$3$\times$3 conv \\stride=2}\\\cline{1-3}

        DRB4 &24$\times$24$\times$4$\times$512& \tabincell{c}{1$\times$1$\times$1 conv \\3$\times$3$\times$3 conv \\num=8}\\\cline{1-3}

        TransLayer4 & 24$\times$24$\times$4$\times$256 & 1$\times$1$\times$1 conv\\\cline{1-3}

        Deconvolution2 &48$\times$48$\times$8$\times$128&\tabincell{c}{ 3$\times$3$\times$3 conv \\stride=2}\\\cline{1-3}

        DRB5 &48$\times$48$\times$8$\times$256&\tabincell{c}{ 1$\times$1$\times$1 conv \\3$\times$3$\times$3 conv \\num=4}\\\cline{1-3}

        TransLayer5 & 48$\times$48$\times$8$\times$128 & 1$\times$1$\times$1 conv\\\cline{1-3}

        Deconvolution3 &96$\times$96$\times$16$\times$64&\tabincell{c}{ 3$\times$3$\times$3 conv \\stride=2}\\\cline{1-3}

        DRB6 &96$\times$96$\times$16$\times$128& \tabincell{c}{1$\times$1$\times$1 conv \\3$\times$3$\times$3 conv \\num=2}\\\cline{1-3}

        TransLayer6 & 96$\times$96$\times$16$\times$64 & 1$\times$1$\times$1 conv\\\cline{1-3}

        Convolution2&96$\times$96$\times$16$\times$1&1$\times$1$\times$1 conv\\\cline{1-3}

        \hline
        \hline
    \end{tabular}
    \label{tab:structure}
\end{table}
\subsection{Densely-connected Residual Block (DRB)}

Let ${x_l}$ be the output of the ${l^{th}}$ convolutional layer, which can be considered as the result of applying a non-liner transformation $H_l$ defined as a convolution followed by a batch-normalization and a rectified linear unit (ReLU) in the ${l^{th}}$ layer. And ${x_0}$ denotes the input data sample passed to the CNN.
For a classical CNN layer with straightforward connection, ${x_l}$ can be modelled as
\begin{equation}\label{eq1}
{x_l} = {H_l}({x_{l - 1}}),
\end{equation}
where $x_{l - 1}$ is the output of the ${(l - 1)^{th}}$ layer.
However, when a network goes deeper, the network suffers from the degradation problem - the gradient may vanish or explode. This phenomenon leads to large training errors and the network training may not converge.

To alleviate the problem by promoting information propagation within the network, in this paper, we propose a new block by combining dense block\cite{huang2017densely} with residual connection as show in Fig.~\ref{fig:network_structure}(B).
The dense connected layers provide a directly connects with all subsequent layers. The feature maps produced by all the preceding layers are concatenated as input for the subsequent layers. Consequently, the ${l^{th}}$ layer receives all feature maps produced by $[0,1,...,l - 1]$ layers as inputs. The output of the ${l^{th}}$ layer is then defined as
\begin{equation}
\label{eq2}
{x_l} = {H_l}([{x_0},{x_1},...,{x_{l - 1}}]),
\end{equation}
where $[{x_0}, {x_1}, \ldots, {x_{l - 1}}]$ represent the concatenation of the feature maps.

To reduce the number of features and efficiently fuse the features from dense layers, a transition layer is added at the end of each dense block. The transition layer consists of a 1$\times$1 convolution layer, a batch-normalization and a ReLU. The out of the transition layer is
\begin{equation}\label{eq3}
{x_t} = {H_t}({H_l}([{x_0},{x_1},...,{x_{l - 1}}])),
\end{equation}
where ${H_t}$ is a non-liner transformation of transition layer.
To further promote information propagation and make the network easier to optimize, we also employ residual connection into our block. 

\subsection{Attention-focused Module (AM)}

To make the network focus more on the region to be segmented and to reduce noise features from the surrounding region, we introduce an attention-focused module in our model.
The structure of attention-focused module is shown in Fig.~\ref{fig:network_structure}(C), which consists of a sigmoid layer and an element-wise multiplication layer.
The output of AM is the element-wise multiplication of input feature-maps and attention masks. The attention masks are produced by sigmoid layer:
 \begin{equation}\label{eq6}
 {M_t}(x) = f({H_t}(x))
 \end{equation}
 \begin{equation}\label{eq7}
 f(x) = \frac{1}{1 + e^{-x}}
 \end{equation}
 where ${M_t}(x)$ denotes the attention mask, whose values range in $[0,1]$, ${H_t}(x)$ denotes the feature map from long connection.

\section{Experiments}
\label{sec:experiments}

To evaluate the performance of our proposed model, we applied the developed method on the MICCAI Prostate MR Image Segmentation 2012 Grant Challenge dataset\footnote{https://promise12.grand-challenge.org/}, TCIA Pancreas CT-82\footnote{https://wiki.cancerimagingarchive.net/display/Public/Pancreas-CT} and MICCAI 2017 Liver Tumor Segmentation (LiTS) Challenge dataset\footnote{https://competitions.codalab.org/competitions/17094} for image segmentation. 

\subsection{Implementation Details}

The proposed method is implemented using the open source deep learning library Keras. Our network is trained end-to-end by using the Stochastic Gradient Descent (SGD) optimization method. In the training phase, the learning rate is initially set to 0.0001 and decreased with a weight decay of 10e-6. The momentum is set to 0.9. Experiments are carried out on a NVIDIA GTX 1080ti GPU with 11GB memory.

\begin{figure}
  \centering
  \includegraphics[width=0.7\columnwidth]{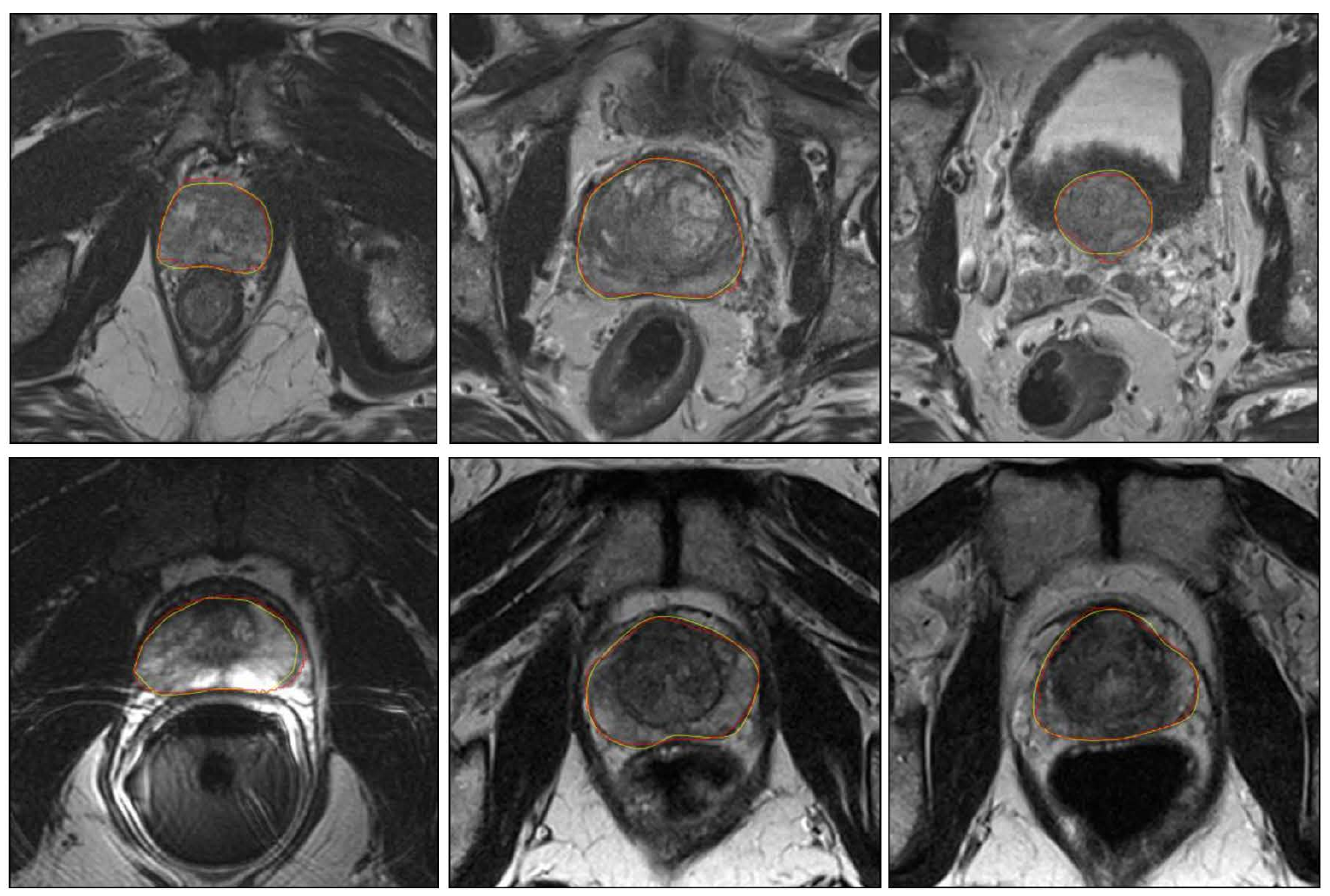}
  \caption{Segmentation results of the prostate from MR images. The yellow and red contours indicate the ground truth and our segmentation results, respectively. Note that these results are directly obtained from challenge website.}\label{FigRE}
\end{figure}

\begin{table}[]
  \caption{Quantitative evaluation results of the proposed method and other methods on prostate MR segmentation.}
  \label{tab:prostate_comparison}
\begin{tabular}{c|c|c|c|c|c|c|c}
\hline \hline
\multicolumn{2}{c|}{User}                                                       & QuIIL & aslm  & GeertLitjens & lanqier xl & tbrosch & {\bf Ours}  \\ \hline         \hline
\multirow{3}{*}{\begin{tabular}[c]{@{}c@{}}ABD\\ {[}mm{]}\end{tabular}}  & Whole & 1.71  & 1.53  & 1.71         & 1.59       & 1.49    & {\bf1.31}  \\ \cline{2-8}
                                                                         & Base  & 1.96  & 1.64  & 1.96         & 1.88       & 1.73    & {\bf1.60}  \\ \cline{2-8}
                                                                         & Apex  & 1.62  & 1.93  & 1.56         & 1.67       & 1.73    & {\bf1.39}  \\ \hline
\multirow{3}{*}{\begin{tabular}[c]{@{}c@{}}HD\\ {[}mm{]}\end{tabular}}   & Whole & 4.92  & 4.62  & 5.13         & 4.63       & 4.68    & {\bf3.97}  \\ \cline{2-8}
                                                                         & Base  & 5.07  & {\bf 4.34}  & 5.22         & 5.22       & 4.90    & 4.75  \\ \cline{2-8}
                                                                         & Apex  & 3.97  & 5.16  & 4.17         & 4.26       & 4.49    & {\bf3.70}  \\ \hline
\multirow{3}{*}{\begin{tabular}[c]{@{}c@{}}DSC\\ {[}\%{]}\end{tabular}}  & Whole & 89.02 & 90.24 & 89.43        & 89.69      & 90.46   & {\bf91.42} \\ \cline{2-8}
                                                                         & Base  & 86.04 & 88.98 & 86.42        & 86.79      & 88.51   & {\bf89.41} \\ \cline{2-8}
                                                                         & Apex  & 86.39 & 83.31 & 86.81        & 86.79      & 85.29   & {\bf88.51} \\ \hline
\multirow{3}{*}{\begin{tabular}[c]{@{}c@{}}aRVD\\ {[}\%{]}\end{tabular}} & Whole & 7.26  & 7.98  &  6.95        & 7.58       & {\bf 6.59}    & 6.97  \\ \cline{2-8}
                                                                         & Base  & 13.57 & 12.68 & 11.04        & 11.63      & 9.64    & {\bf8.53}  \\ \cline{2-8}
                                                                         & Apex  & 16.70 & 18.92 & 15.18        & 14.92      & 18.51   & {\bf13.05} \\ \hline
\multicolumn{2}{c|}{{\bf Overall score}}                                              & 86.71 & 86.89 & 87.15        & 87.21      & 87.67   & {\bf89.18} \\ \hline \hline
\end{tabular}
\end{table}

\subsection{Prostate Segmentation From MR Image}
We first evaluated our proposed method on MICCAI 2012 Prostate MR Image Segmentation (PROMISE12) challenge dataset. There are in total 50 transversal T2-weighted MR images of the prostate and the corresponding ground truth segmentation, which were checked and corrected by a radiological resident with more than 6 years of experience in prostate MRI. These images are a representative set of the types of MR images acquired in different hospitals. And these images are from multiple vendors and have different acquisition protocols and variations in voxel size, dynamic range, position, field of view and anatomic appearance. To evaluate the proposed algorithms, the organizers provide 30 testing MR images and the corresponding ground truth is held out.

Before training the network, we resampled all MR volumes into a fixed resolution of 0.625$\times$0.625$\times$1.5mm and then normalized them as zero mean and unit variance. To facilitate network training, we applied data augmentation operations including rotation, scaling and flipping. During training, we adopted a random cropping strategy, where sub-volumes in the size of 16$\times$96$\times$96 ($d\times w \times h$) voxels are randomly cropped from the training data during every iteration. In the testing phase, similar to the works in \cite{yu2017volumetric,yu2017automatic}, we used overlapping sliding windows to crop sub-volumes and used the average of the probability maps of these sub-volumes to get the whole volume prediction. The sub-volume size was also 16$\times$96$\times$96 and the stride was 8$\times$48$\times$48. Due to the limitation of the memory, we used the mini-batch size of 4. The number of parameters of SIP-Net was 3.16M, and the prediction time was approximately 1 minutes for one MR volume.

Several sample results of our proposed method are shown in Fig.~\ref{FigRE}. It can be seen that our model can accurately segment the prostate and obtain smooth and continuous prostate boundaries.
Quantitative evaluation was also performed.
The evaluation metrics used in PROMISE12 challenge include Dice Similarity Coefficient (DSC), percentage of the absolute difference between the volumes (aRVD), average over the shortest distance between the boundary points of the volumes (ABD) and Hausdorff Distance (HD). All the evaluation metrics are calculated in 3D. In addition to evaluating these metrics over the entire prostate segmentation, the challenge organizers also calculated the boundary measures specifically for the apex and base parts of the prostate, because those parts are difficult to segment but in the same time very important for many clinical applications.
The apex and base the prostate are determined by dividing the prostate into three approximately equal sized parts along the axial direction (the first 1/3 as apex and the last 1/3 as base).
Then an overall score will be computed by taking all the criteria into consideration rank the algorithms.

The results of our proposed method and the competitors are shown in Table~\ref{tab:prostate_comparison}. Only the top 10 teams are listed. Note that all the results reported in this section were obtained directly from the challenge website. As it can be seen from the table, our overall performance was the best and therefore ranked the first place among all the teams (by May 22, 2018)\footnote{https://promise12.grand-challenge.org/evaluation/results/} with the score of 89.18.
From Table~\ref{tab:prostate_comparison}, it can be seen that our proposed model achieved the best performance in several measures.
The segmentation results of our model were the best not only for whole prostate segmentation, but also in the base and apex areas, which demonstrates the effectiveness of the proposed 3D model with DRBs and AM modulated long connections.


\begin{figure}
  \centering
  \includegraphics[width=0.7\columnwidth]{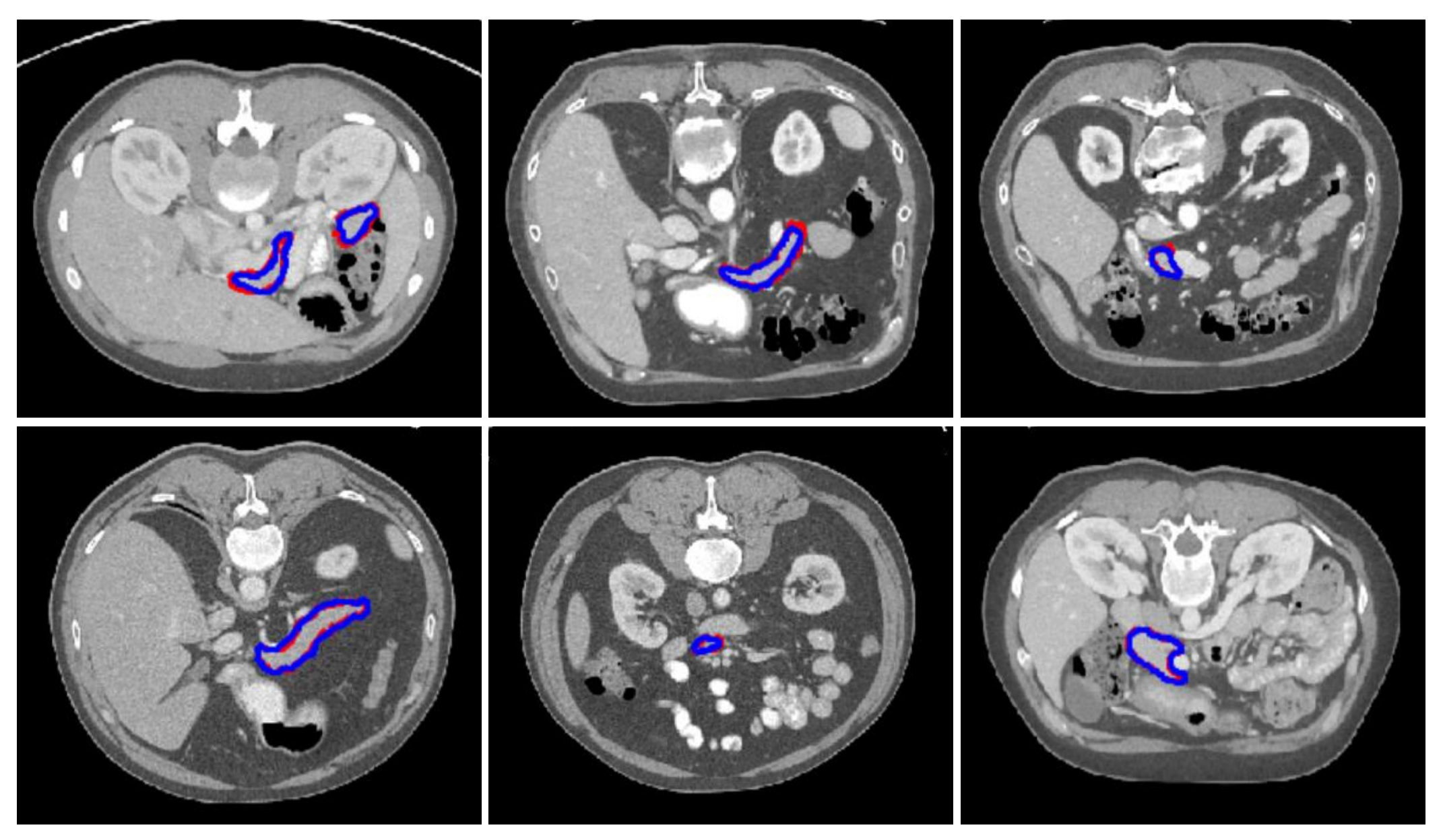}
\caption{Sample segmentation results of the pancreas CT images. The red and blue contours are the ground truth and our segmentation results, respectively.}\label{FigPR}
\end{figure}

\begin{table}
\caption{Performance of CNN based CT pancreas segmentation methods, which are trained and evaluated using the same number of training and testing images.}
  \centering
  \fontsize{8}{12}\selectfont
    \begin{tabular}{l|l}
        \hline         \hline
         Methods & DSC [\%]\\
        \hline         \hline

        Holistically Nested 2D FCN Stage-1 \cite{roth2018spatial} & 76.8 $\pm$ 11.1 \\

        Holistically Nested 2D FCN Stage-2 \cite{roth2018spatial} & 81.2 $\pm$ 7.3\\

        2D FCN \cite{cai2017improving}& 80.3 $\pm$ 9.0\\

        2D FCN + Recurrent Network \cite{cai2017improving}& 82.3 $\pm$ 6.7\\

        Single Model 2D FCN \cite{zhou2017fixed}& 75.7 $\pm$ 10.5\\

        Multi-Model 2D FCN \cite{zhou2017fixed}& 82.2 $\pm$ 5.7\\

        Attention U-Net \cite{oktay2018attention}& 81.5 $\pm$ 6.2\\

        SIP-Net (Our) &\textbf{83.9 $\pm$ 4.5}	\\

        \hline         \hline
    \end{tabular}
\label{tab:pancreas}
\end{table}

\subsection{Pancreas Segmentation}

The proposed model is also evaluated on another publicly available dataset -- TCIA Pancreas CT-82. This dataset contains 82 contrast enhanced 3D CT scans, which have resolutions of 512$\times$512 pixels with varying pixel sizes and slice thickness between 1.5-2.5 mm, acquired on Philips and Siemens MDCT scanners \cite{Data}. The dataset is publicly available and commonly used to benchmark CT pancreas segmentation frameworks. In our experiments, the 82 scans are randomly split with 62 images for training and 20 images for testing. Before training the model, we resampled all volumes into a fixed resolution of 1.0mm$\times$1.0mm$\times$1.0mm. Then all the scans are normalized to have zero mean and unit variance.
We again applied data augmentation operations including rotation, scaling and flipping.
We also employed the random cropping strategy, where sub-volumes in the size of 64$\times$96$\times$96 ($d \times w \times h$) voxels are randomly cropped from the training data during every iteration. In the testing phase, we used overlapping sliding windows to crop sub-volumes and used the average probability maps of these sub-volumes to get the whole volume prediction. The sub-volume size was also 64$\times$96$\times$96 and the stride was 32$\times$48$\times$48. The architecture of network was same as that utilized on prostate segmentation. The prediction time was approximately 1 minutes for one CT volume.

To evaluate the proposed architecture, we compare the performance of the model against other state-of-the-art CT pancreas segmentation methods. The results are summarized in Table~\ref{tab:pancreas}. It can be seen that our proposed model achieved 83.9 $\pm$ 4.51 in DSC for pancreas labels, which outperform other state-of-the-art methods. Several example segmentation results of our proposed method are shown in Fig.~\ref{FigPR}. Our proposed model can accurately segment the pancreas from CT images. It is worth noting that we only employ a single model to segment pancreas and our model does not require multiple CNN models as in~\cite{roth2018spatial}.

\subsection{Liver Segmentation}

We also tested our proposed model on the competitive dataset of MICCAI 2017 LiTS Challenge, which contains 131 contrast enhanced 3D abdominal CT scans with radiologist hand-drawn ground truths for training and the rest 70 used for testing with unreleased ground truth. Since the data were acquired from different clinical sites, which have different scanners and protocols, the scans have largely varying in-plane resolution (0.55mm-1.0mm) and slice spacing (0.45mm-6.0mm). Before training the model, we truncated the image intensity values of all scans to the range of [-200,200] to remove the irrelevant details and then normalized each volume. In addition to 3D model, we also evaluate the 2D model with same network structs for evaluating the influence of parameters.
\begin{table}
\caption{Quantitative evaluation results of the proposed method and other methods on MICCAI 2017 LiTS Challenge Dataset.}
  \centering
  \fontsize{8}{12}\selectfont
    \begin{tabular}{l|c|c}
        \hline         \hline
         Methods & Per Case DSC [\%]  & Global DSC [\%] \\
        \hline         \hline
        H-DenseUNet \cite{li2018h} & 96.1 & 96.5 \\

        CascadedResNet\cite{bi2017automatic} & \textbf{96.3} & \textbf{96.7} \\

        SIP-Net(2D) (Ours) & 95.9 & 96.3 \\
        SIP-Net(3D) (Ours) & 94.2 & 94.6 \\
        \hline         \hline
    \end{tabular}
\label{tab:liver_comparison}
\end{table}

During the network training, we randomly cropped patches in the size of 224$\times$224$\times$16 pixels for 3D model (224$\times$224 pixels for 2D model) from the training data during every iteration. In the testing phase, we used overlapping sliding windows to crop sub-volumes and used the average probability maps of these sub-volumes to get the whole volume prediction. The cropped size was also 224$\times$224$\times$16 pixels for 3D model (224$\times$224 pixels for 2D model) and the stride was 112$\times$112$\times$8 for 3D model (112$\times$112 for 2D model).
The number of parameters of SIP-Net (2D) was 1.43M, and the prediction time was approximately 2 minutes for one CT volume.

There were more than 60 submissions for the MICCAI LiTS Challenge. The segmentation performances of the teams are listed on the leaderboard\footnote{https://competitions.codalab.org/competitions/17094\#results} and we were among the top seven teams (by November 15, 2018, team of Qikui\_sigma-RPI). We compared the performance of our model with two published top-performance models: H-DenseUNet \cite{li2018h} and CascadedResNet \cite{bi2017automatic}. H-DenseUNet employed a simple ResNet to process the original data, which makes the network subject to the performance of pro-processing. In addition, H-DenseUNet employed 3D convolutional layers inside the model with much more parameters and thus increased training difficulty. CascadedResNet, on the other hand, achieved good results but took approximately 7 days on two Titan X GPUs for training. Our proposed method performs similarly to the above two approaches with negligible differences, however, can be trained much more efficiently than those methods. Comparing the performance of 2D and 3D models reveals that the 2D model can even obtain better performance. This indicates that the network architecture is the key for the performance gain and a larger number of network parameters may lead to performance decrease.

\section{Discussions}
\label{sec:discussions}

In this section, we provide in-depth discussions of the effects of some of our proposed components.

\subsection{Ablation Study of Network Structure}

In order to evaluate the effectiveness of the residual connections in dense blocks, the long connections and attention-focused modules used in our model, we performed a set of ablation study experiments.
The prostate MR image dataset was used. We randomly selected 10 patients for validation and the rest 40 patients were utilized for training.
\begin{table}
  \caption{Performance of the proposed model in different configurations.}
  \centering
  \fontsize{8}{12}\selectfont
    \begin{tabular}{l|c}
        \hline         \hline
         Configurations & Global DSC [\%]\\
        \hline         \hline
        D-Net & 86.0 \\
        DR-Net & 86.9\\
        DRL-Net & 88.8\\
        SIP-Net (Ours) & \textbf{89.8}	\\
        \hline         \hline
    \end{tabular}
    \label{tab:ablation}
\end{table}

\begin{figure}
  \centering
  \includegraphics[width=0.7\columnwidth]{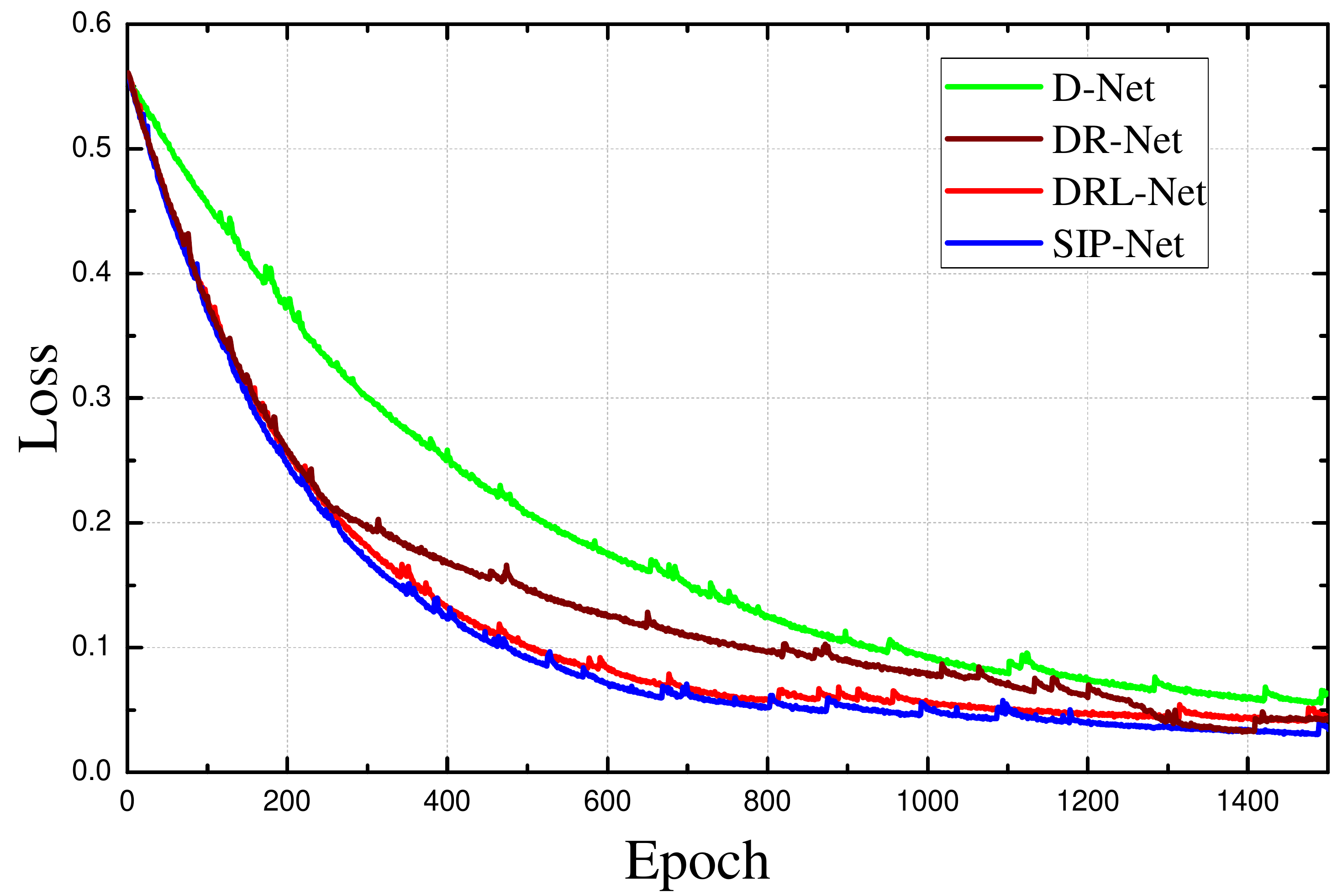}
  \caption{Training loss of networks with different structures.}\label{FigLoss}
\end{figure}
\begin{figure}
  \centering
  \includegraphics[width=0.7\columnwidth]{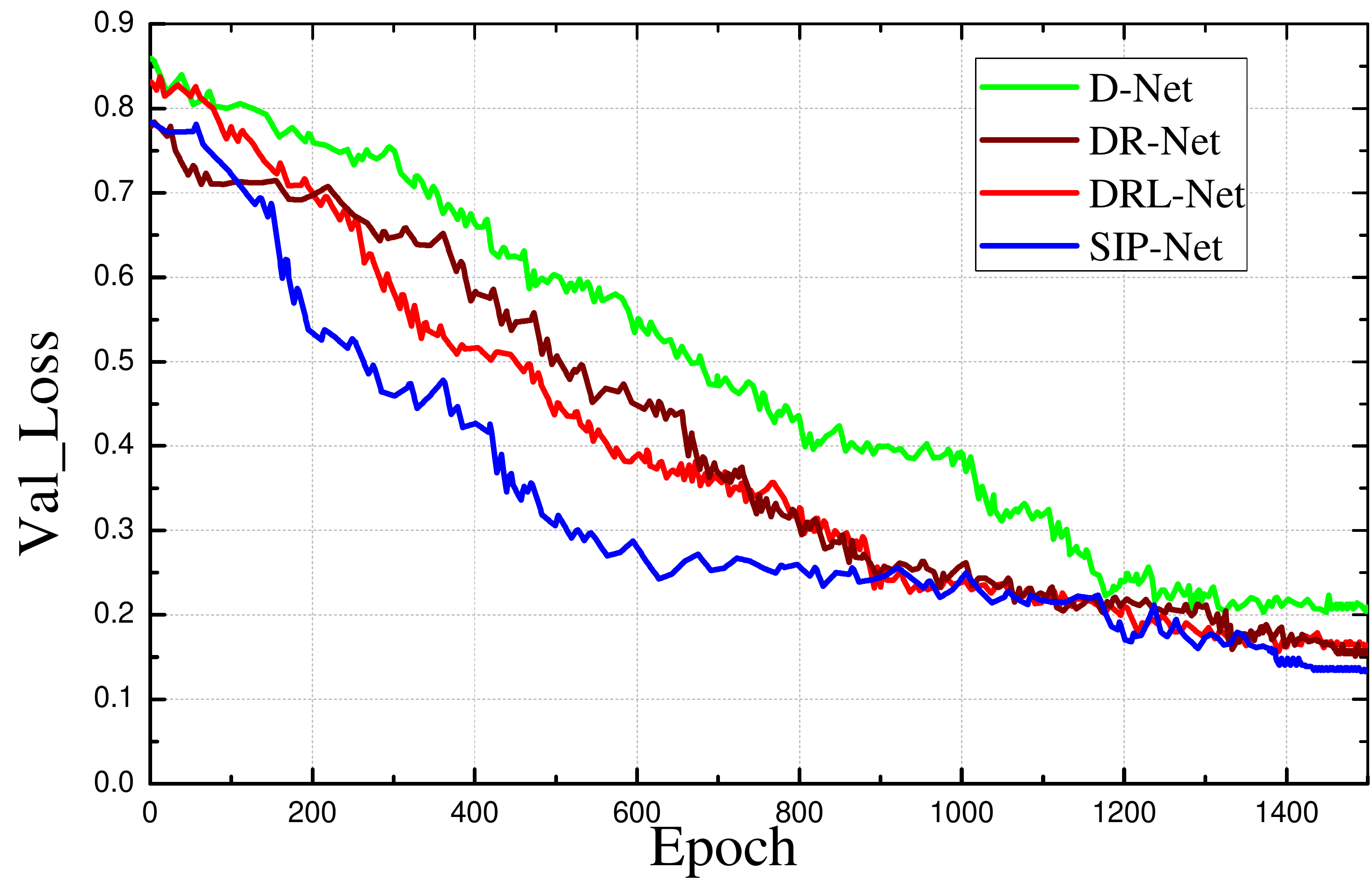}
  \caption{Validation loss of the networks with different structures.}\label{FigVal}
\end{figure}

To analyze the learning behaviors of our model, we created four different configurations of our model: using only dense block (D-Net), using only DRBs (DR-Net), using DRBs and long connections (DRL-Net), using DRBs, long connections and attention-focused module (SIP-Net). We first analyzed the leaning behaviors of these models. Figs.~\ref{FigLoss} and \ref{FigVal} present the training and validation losses of different networks. It is observed that the models with either residual connections, long connections and attention-focused module converge faster and achieve lower validation loss than the one with only dense block, which demonstrates that the use of residual connections, long connections and attention-focused modules can improve the training efficiency and the performance of the models. Fig.~\ref{FigVal} further shows that the long connections can accelerate the convergence speed and alleviate the risk of over-fitting on limited training data.

Table~\ref{tab:ablation} shows the performance of our proposed model with different connections and blocks. It is can be seen that adding residual connections, long connections and attention-focused modules can achieve better Dice scores than the network with only dense blocks. The network with residual connections and dense block has marginally better performance than that with only dense block, which demonstrates that the enhanced information propagation inside each block can improve the performance of the model. The model with long connections obtained better performance than the one without. It is conceivable that enhancing information propagation both locally and globally inside the model and combining them together can further improve the performance. The network with attention-focused modules achieves the best performance in the ablation experiments, indicating that attention-focused module further improves the performance of model.

\begin{figure}
  \centering
  \includegraphics[width=0.7\columnwidth]{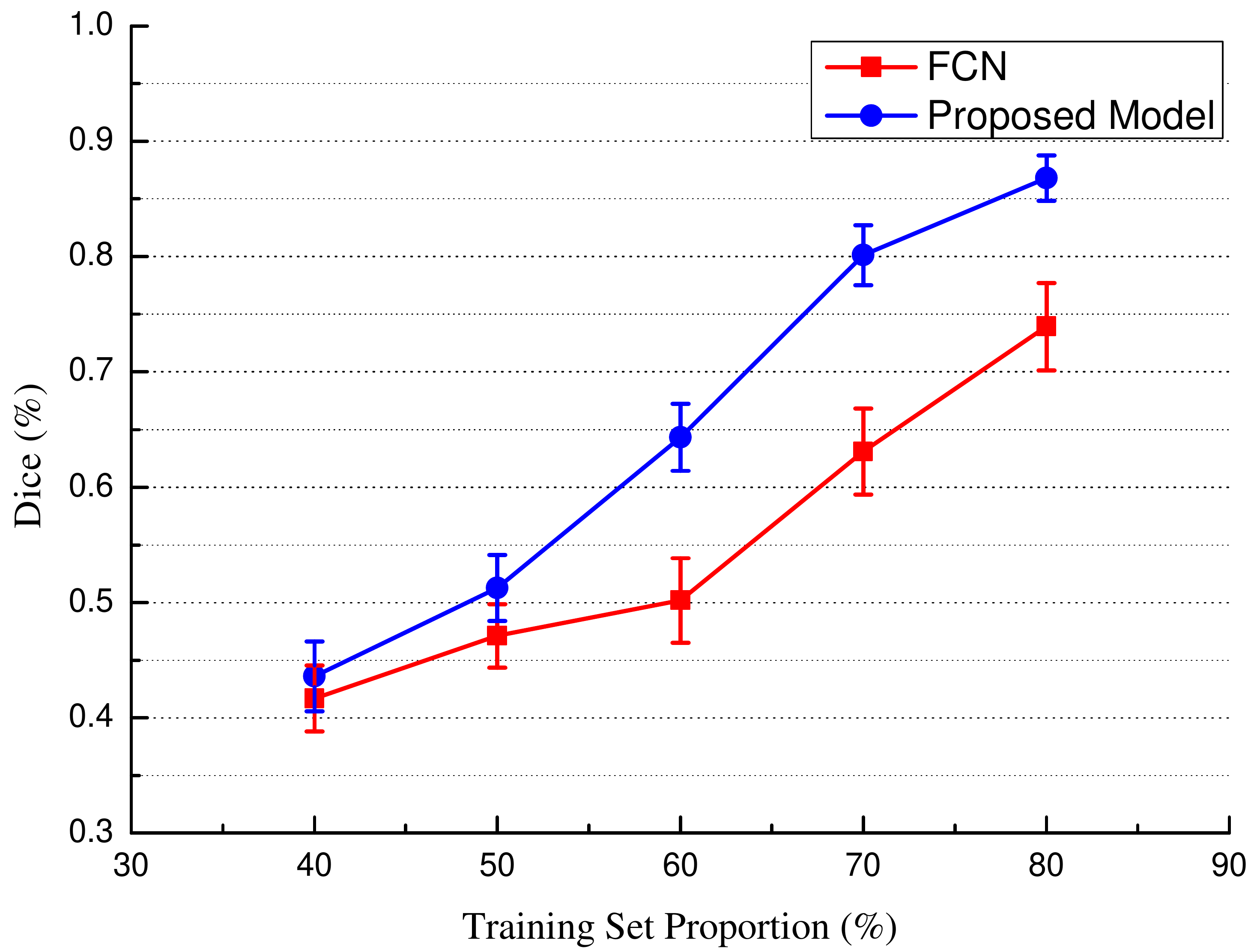}
  \caption{Performance of the proposed model and FCN under different training set proportions.}\label{FigCompareResults}
\end{figure}

To demonstrate the efficiency of the proposed method in utilizing training data, we compare the performance of the model against that of FCN, which is indeed the version of our model without DRB and AM, using different amount of training data. In this experiment, we respectively used 40\%, 50\%, 60\%, 70\%, and 80\% of data for training and reserved upto 20\% of the data for testing. To avoid potential data distribution bias, in each setting, we randomly selected five different subsets from the entire dataset for training and testing. The average performances over the five runs under each setting are reported and shown in Fig.~\ref{FigCompareResults}. It can be seen that, when only 40\% of the training data were used, the proposed method and FCN achieved similar performance. The performances are poor as the training data is very limited in that case as we expected. As the size of the training dataset increases, both methods start to perform better. However, Fig.~\ref{FigCompareResults} shows that the proposed method improves in a much faster rate, with the contribution from the proposed DRB and AM modules. Eventually, the proposed model only needs less than 70\% of the training data to outperform the FCN trained with the entire 80\% of the data. The experiment demonstrates that the proposed structures can help deep CNNs get trained more efficiently with small number of training images, which is a very desired property for medical imaging applications where labeled data is usually in scarce.

\begin{figure}
  \centering
\includegraphics[width=0.5\columnwidth]{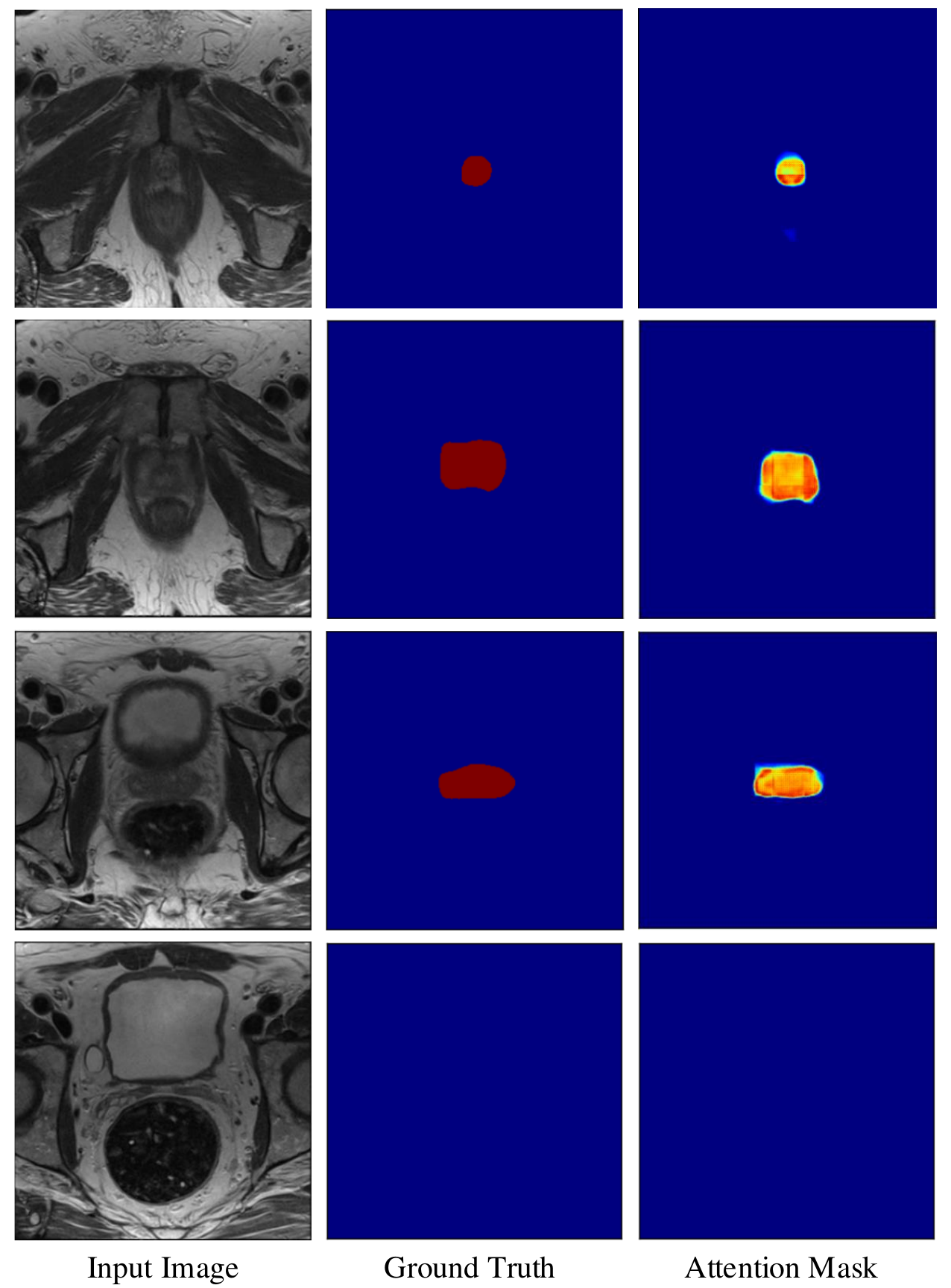}
  \caption{Attention mask examples produced by the attention-focused module. The blue, red pixel represents background and prostate, respectively. And the attention mask is the corresponding heat map produced by attention-focused modules. For the heat map, the darker the color, the greater the weight value, the lighter the color, the smaller the weight value.}

\label{FigAM}
\end{figure}

\subsection{Analysis of Attention-focused Modules}

To further analyze the function of attention-focused modules, we visualized the generated attention masks in the up-sampling path. Four different types of input images were selected, which are selected from base, mid-gland, apex and also outside of the prostate. It can be seen that the attention masks have much higher weights in the prostate region than in the non-prostate region as shown in Fig.~\ref{FigAM}. And the shape of attention mask was very close to the ground truth.  It is conceivable that higher weight was inside the attention masks, which helps to locate the region of prostate. The shape of the attention mask volume was again close to the ground truth. It suggests that the attention mask can help the network pay more attention to the region of prostate and suppress the features from the non-prostate region towards better image segmentation.

\subsection{Effects of Batch Size}
To evaluate the influence of batch size on the segmentation results, we compared the performance of our proposed model under various batch size. The prostate MR image dataset also was used, 10 patients were randomly selected for validation and the rest 40 patients were utilized for training. The segmentation performance is listed in Table~\ref{tab:bs}. It can be seen that the size of batch has a slight effect on the segmentation results and the model performed the best when batch size is 4.
\begin{table}[]
 \caption{Performance of SIP-Net in different batch size.}
  \centering
  \fontsize{8}{12}\selectfont
\begin{tabular}{c|c|c|c|c}
\hline \hline
Batch-size & 1 & 2 & 3 & 4 \\ \hline
DSC [\%]   & 89.4 & 89.5 & 89.7 & 89.8  \\ \hline \hline
\end{tabular}
    \label{tab:bs}
\end{table}

\section{Conclusions}
\label{sec:conclusions}

In this paper,  we first prove that not all the feature maps transmitted by skip connections contribute positively to the network performance. And to adaptive select information passed through those skip connections, we propose a novel network, named SIP-Net, which can adaptive select the information passed through those skip connections by our proposed attention-focused modules. Expect for making the skip connections between the down-sampling path and up-sampling path can further improve the context and gradient information propagation both forward and backward and address the vanishing-gradient problem, our proposed SIP-Net also makes the model focus on the region of interest.
Extensive experiments on the publicly available MICCAI Prostate MR Image Segmentation 2012 Grant Challenge dataset, TCIA Pancreas CT-82 and MICCAI 2017 Liver Tumor Segmentation (LiTS) Challenge dataset demonstrate that our proposed method can get more accurate  boundaries  and  achieve  superior  results  compared with  other  state-of-the-art  methods.



%
 \section*{Conflict of interest}
 The authors declare that they have no conflict of interest.

\bibliographystyle{unsrt}        
\bibliography{mybibfile}
%
%

\end{document}